\documentclass[aps,prd,twocolumn,amsmath,amssymb,showpacs,floatfix,nofootinbib]{revtex4-1}
\usepackage[usenames]{color}
\usepackage[normalem]{ulem}
\usepackage{amssymb,amsmath,graphicx,color,microtype}
\usepackage{graphicx}
\usepackage{txfonts}
\usepackage{epsf}
\usepackage{epstopdf}
\usepackage {amssymb}
\usepackage{multirow}
\newcommand{\nc}{\newcommand}
\nc{\ba}{\begin{eqnarray}}
\nc{\ea}{\end{eqnarray}}
\newcommand\be{\begin{equation}}
\newcommand\ee{\end{equation}}

\nc{\e}{{\bf{e}}}
\nc{\kk}{{\bf{k}}}
\nc{\pp}{{\bf{p}}}

\nc{\bfk}{{\bf{k}}}
\nc{\bfx}{{\bf{x}}}
\nc{\bfp}{{\bf{p}}}

\nc{\eH}{{\epsilon_H}}
\nc{\calP}{{\cal P}}
\nc{\im}{{ \mathrm{Im} } }

\begin{document}
\title{Formation Redshift of the Massive Black Holes Detected by LIGO}

\author{Razieh Emami}
\email{razieh.emami$_$meibody@cfa.harvard.edu}
\author{Abraham~Loeb}
\email{aloeb@cfa.harvard.edu}
\affiliation{Institute for Theory and Computation, Harvard University, 60 Garden Street, Cambridge, MA 02138, USA}

\begin{abstract}
We compare the event rate density detected by LIGO to the comoving number density of isolated stellar progenitors and find a range for their formation redshift. 
Our limit depends on the threshold mass for making the black holes (BHs) but only weakly on the metallicity of their progenitor. If $10 \%$ of all BHs are in coalescing binaries, then enough progenitors have formed by $  2< z_f < 9 $. 
\end{abstract}

\maketitle
The detection of the gravitational waves (GWs) by the Laser Interferometer Gravitational Wave Observatory (LIGO) \cite{TheLIGOScientific:2016pea,TheLIGOScientific:2016qqj,Abbott:2016nhf,Abbott:2017vtc,Abbott:2017gyy,Abbott:2017oio}, from the merger of binary black holes (BBHs) ushered a new era of multi-messenger astronomy. In this \textit{Letter} we focus on the events with BH masses above $20 M_{\odot}$ which are according to most recent data from the 
LIGO/Virgo Collaboration (LVC) \cite{LIGOScientific:2018jsj, LIGOScientific:2018mvr} are about 
70 \% of the detected events.

The LIGO results prompted two theoretical challenges. One involves the large masses of the detected BHs and the second involves their assembly into binaries that coalesce within the age of the universe. There are three main channels proposed to address these issues.. ``\textit{Dynamical formation}" requires a dense star cluster. In this channel, BHs are formed through the evolution of massive stars and segregate to the cluster core where they pair as BBHs \cite{Rodriguez:2015oxa,Rodriguez:2016kxx,Park:2017zgj}. ``\textit{Classical isolated binary evolution}" leads to the formation of a BBH through a common envelope ejection of an expanded envelope beyond its Roche Lobe \cite{Belczynski:2014iua,Belczynski:2016obo,Woosley:2016nnw,Rodriguez:2018rmd,Choksi:2018jnq}. 
In ``\textit{chemically homogeneous evolution}", mixing plays  an important role in spreading the helium produced at the core throughout the envelope and causing an almost homogeneous evolution of the progenitor stars to BHs \cite{deMink:2010zm,deMink:2016vkw}.
Alternative formation channels may also be possible 
\cite{Loeb:2016fzn, DOrazio:2017bld}.

Supposing that BBHs are the remnants of gravitationally collapsed progenitors and ignoring the possibility of primordial BHs \cite{Bird:2016dcv}, it is important to establish the connection between their mass and their progenitor mass at zero age on the main sequence. Following the process of gravitational collapse, the resulting BH mass depends not only on the progenitor mass but also on other parameters, including the metallicity, rotation and magnetic field. Therefore, instead of  a one-to-one match between the progenitor and the BH mass, we consider a range of progenitor masses for each BH mass \cite{Woosley:2014lua, Sukhbold:2017cnt,Heger:2002by} as follows:
\begin{itemize}
\item $\mathbf{\leq 30 M_{\odot}}$:
Collapse to neutron stars or light BHs. Since our discussion focuses on massive BHs, we ignore this range.
\item $\mathbf{30-80~M_{\odot}}$:
Collapse to a BH. The mass of remnant depends on the metallicity of star. 
Metal poor stars with a main sequence mass above $50 M_{\odot}$, could lead to a BH with mass about $20 M_{\odot}$. On the other hand, single stars with very low metalicity, $ Z \lesssim 10^{-3} Z_{\odot}$, barely lose mass and so the remnant mass is close to their original mass \cite{Spera:2017fyx}. Likewise in the chemically homogeneously evolving stars in very metal poor environment  the entire star can be turned to BHs with large masses \cite{Marchant:2016wow}. 

\item$\mathbf{80-150 M_{\odot}}$:
Pulsational pair-instability supernovae (PPSN). The mass loss during the pulsation depends on metallicity but allows BH remnants with a mass above $30 M_{\odot}$ as mentioned in \cite{Woosley:2016hmi}.
\item$\mathbf{\geq 150 M_{\odot}}$:
Stars with a mass between $150-260 M_{\odot}$ yield pair-instability supernovae (PISN) with no remnant. Stars with masses above $260 M_{\odot}$ collapse to heavy BHs. Since LIGO did not observe BH masses above $40 M_{\odot}$ \cite{Fishbach:2017zga}, we neglect all progenitor masses above $150 M_{\odot}$ in our analysis.
\end{itemize}
Hereafter, we consider stars with main sequence masses in the range of $50-150 M_{\odot}$ as progenitors of BHs with mass $\gtrsim 20 M_{\odot}$. 
We use the most recent data from LIGO Virgo Collaboration (LVC) for the merger rate density of the BBH \cite{LIGOScientific:2018jsj, LIGOScientific:2018mvr} given by 
$\mathcal{R}_L = 64.9 ^{+ 75.5}_{-33.6} \rm{Gpc}^{-3} yr^{-1}$.  This range is associated with the flat merger rate, constant in time.  In addition, the LVC found an upper mass for each of BHs in  BBH as $M_{max} = 41.6 ^{+ 9.0}_{-4.5} M_{\odot}$, with a lower mass limit $M^{L}_{min} = 5.0 M_{\odot}$. Here, we wish to find the merger rate density for the most massive BHs with $m \geq 20.0 M_{\odot}$, hereafter $\mathcal{R}_{20}$. We compare this rate with the cosmological star formation density (SFD) and find a range for the formation  redshift happens when these two numbers match.The results depends on various parameters, including the BH formation efficiency from massive star progenitors as well as the fraction of BHs that reside in binaries which coalesce within the age of the universe. 

We start by computing the merger rate density for the populations of BHs with  masses in the range $ M_{min} \leq m \leq M_{max}$. As mentioned above, for massive BHs we may adopt $M_{min} = 20 M_{\odot}$. We notice that this is different than the LIGO lower limit $M^{L}_{min} = 5 M_{\odot}$. Here, a super script $L$ refers to the LIGO choice.

More precisely, we wish to find which fraction of the global event rate density detected by LIGO, hereafter $\mathcal{R}_L$,\cite{LIGOScientific:2018jsj, LIGOScientific:2018mvr}, originates from BBHs above some mass range. For this purpose, we use the standard expression for $\mathcal{R}_L$ \cite{Abbott:2016nhf, Fishbach:2017zga}, 
\ba 
\label{LIGO-Event-Rate-density}
\mathcal{R}_L  \equiv \frac{\Lambda_L}{\langle VT\rangle|_L},
\ea
where $\Lambda_L$ denotes the \textit{expected} number of the BBHs, and
$\langle VT\rangle|_L$ refers to the population-averaged spacetime volume \cite{Abbott:2016nhf, Fishbach:2017zga},
\ba 
\label{averaged-VT}
\langle VT\rangle|_L &=& \int_{M_{min}}^{M_{max}}dm_1 \int_{M_{min}}^{m_{1}} dm_2 VT(m_1,m_2) p_{pop}(m_1,m_2), \nonumber\\
\ea
with the outer integral taken over $ M_{min}  \leq m_1  \leq M_{max}$. Hereafter we use the above values for $M_{min}$ and $M_{max}$. 

$VT(m_1,m_2)$ represents the spacetime volume in which LIGO can detect the binaries, based on the search time and detector sensitivity. It has been shown  \cite{Fishbach:2017zga} that $VT \propto m_1^{k}$ with $ k \sim 2.2$. In Eq. (\ref{averaged-VT}), $p_{pop}(m_1,m_2)$ denotes the mass distribution of BBHs \cite{LIGOScientific:2018jsj, LIGOScientific:2018mvr}. We focus on the following power-low choice  of the mass distribution, as adopted by the LVC. This enables us to use the most recent observational results and convert them to our mass limit.
\ba  
\label{distribution}
p_{pop}(m_1, m_2) \propto \frac{m_1^{-\alpha}}{m_1- M^{L}_{min}} ~~,~~ \alpha = 0.4^{+1.3}_{-1.9}.
\ea
In the following, we use the above form of  $p_{pop}(m_1, m_2)$ with $M^{L}_{min} = 5 M_{\odot}$ and we only allow the normalization vary from the original mass distribution. Using the new form of the mass distribution, we compute  $\langle VT(m_1,m_2) \rangle$ and corresponding $\mathcal{R}_{20}$. In addition, we use the most recent power-low index, $\alpha$, from  \cite{LIGOScientific:2018jsj, LIGOScientific:2018mvr}.

Therefore new LIGO event rate is given by,  
\ba  
\label{min-R}
\mathcal{R}_{20} &=& \frac{\Lambda_{20}}{\langle VT\rangle |_{20}} \nonumber \\
&=& \mathcal{N}_{20} \left(\frac{\langle VT\rangle |_{L}}{\langle VT\rangle |_{20}}\right) \mathcal{R}_L,
\ea
where $\mathcal{N}_{20}$ refers to the ratio between the \textit{expected} LIGO rate for BH masses above some threshold and the total expected rates. In our analysis, we adopt the latest results from \cite{LIGOScientific:2018jsj, LIGOScientific:2018mvr} which imply that about 70\% of the events are associated with $m \geq 20 M_{\odot}$. . This yields $\mathcal{N}_{20} = 7/10$.

Plugging the above numbers in Eq. (\ref{min-R}) yields,
\ba  
\label{caseAB-rate}
\mathcal{R}_{20} &=& \left( \frac{7}{10} \right) \left(\frac{\int_{M^{L}_{min}}^{M_{max}}   m^{2.2-\alpha} dm} {\int_{M^{L}_{min}}^{M_{max}}   m^{2.2-\alpha} dm} \right)  \left[
\frac{\int_{M_{min}}^{M_{max}} m^{2.2- \alpha} 
\left( \frac{m - M_{min}}{m - M^L_{min}} \right) }{\int_{M_{min}}^{M_{max}} m^{- \alpha} \left(  \frac{m - M_{min}}{m - M^L_{min}} \right) }
\right]^{-1} \mathcal{R}_L.
\nonumber\\
&=& \left( 21.67 ^{+ 28.02}_{-20.99} \right) \rm{Gpc}^{-3} yr^{-1}. 
\ea

Next, we compute  the mass density, $\rho_{20}$, of LIGO BBHs progenitors with masses above $20 M_{\odot}$. We assume that each component of the binary originated from the collapse of a star with a zero age main sequence mass  above a threshold mass, hereafter $M_{\star,min}$, which we take to be in the range $ 50 M_{\odot} \leq M_{\star,min} \leq 150 M_{\odot}$. Furthermore since the BH mass is only a fraction of the main sequence star, we use a simple mapping between the stars on the progenitor star mass and the BH mass. Throughout our analysis, we use the mapping of Ref. \cite{Belczynski:2016obo} in their figure 5 for two different metallicities $Z = 0.5 \% Z_{\odot} $ and $Z = 10 \% Z_{\odot} $. 
Each binary system requires two stellar progenitors. For the initial mass function of progenitor stars, we adopt the Kroupa form \cite{Kroupa:2002ky}, $\Phi(M_{\star})$. We integrate $\mathcal{R}(t)$ over cosmic time and take into account the lower mass limit for the progenitor mass, $M_{\star,min}$. The required comoving mass density of progenitor stars is therefore,
\ba 
\label{LIGO-Number-density}
\rho_{20}(M_{\star,min}) &=& 2 \left(\int_{0}^{t_{H}} \mathcal{R}_{20}(t)~ dt \right) \times \frac{\int_{M_{\star,min}}^{150} M_{BH}(M_{\star}) \Phi( M_{\star}) d M_{\star} }{\int_{M_{\star,min}}^{150} \Phi( M_{\star}) d M_{\star} } \nonumber\\
&\simeq& 2  \mathcal{R}_{20} ~ t_{H} \times \bigg(\frac{\int_{M_{\star,min}}^{150} M_{BH}(M_{\star}) \Phi( M_{\star}) d M_{\star} }{\int_{M_{\star,min}}^{150} \Phi( M_{\star}) d M_{\star} }  \bigg).
\ea
Here we use one of the most recent models used  by the LVC with a constant merger rate in time \cite{LIGOScientific:2018jsj, LIGOScientific:2018mvr}. It is satisfactory to generalize this study to the more complicated case with a power-low redshift dependence as well as a time delay in the BBH formation, which goes beyond the limited scope of this paper. Thus we use the 
average value of $\mathcal{R}(t)$ over the age of the universe $t_{H} = 1.38 \times 10^{10} \rm{yr}$ to be close to the estimated value by LIGO at 
$z  \simeq 0.18$ \cite{Abbott:2017vtc}.

Next, we compute the star formation density $\rho_{\star}$ for some fraction of the stars above a threshold, $ 50 M_{\odot} \leq M_{\star,min} \leq 150 M_{\odot}$. 
For this purpose, we adopt the star formation rate density (SFRD) as presented in \cite{Madau:2016jbv} for 
$z\leq 8$ 
and in \cite{Finkelstein} for $z \geq 8$.
The uncertainty in the SFRD at $z\gtrsim 8$ has a very weak effect on our results. 

We integrate the SFRD over cosmic time to get the global star formation density, as inferred from the observed UV luminosity density in the universe as a function of redshift. Since the UV emission is dominated by massive stars, we do not expect our results to be very sensitive to the assumed form of $\Phi(M_\star)$.
As noted above, we need to make sure that the remnants of the stars are within our desired mass range for LIGO's BBHs. This can be done by multiplying the $\rho_{\star}$ with the factor,
\ba 
\label{fmin}
f_{min}(M_{\star,min})&\equiv& \left(\int_{M_{\star,min}}^{150} M_{\star} \Phi(M_{\star}) dM_{\star} \right) \bigg{/} \left( \int_{0}^{150} M_{\star}\Phi(M_{\star}) dM_{\star} \right), \nonumber\\
\ea
where we consider $ 50 M_{\odot} \leq M_{\star,min} \leq 150 M_{\odot}$.  

\begin{figure*}
 \center
   \includegraphics[width=\textwidth]{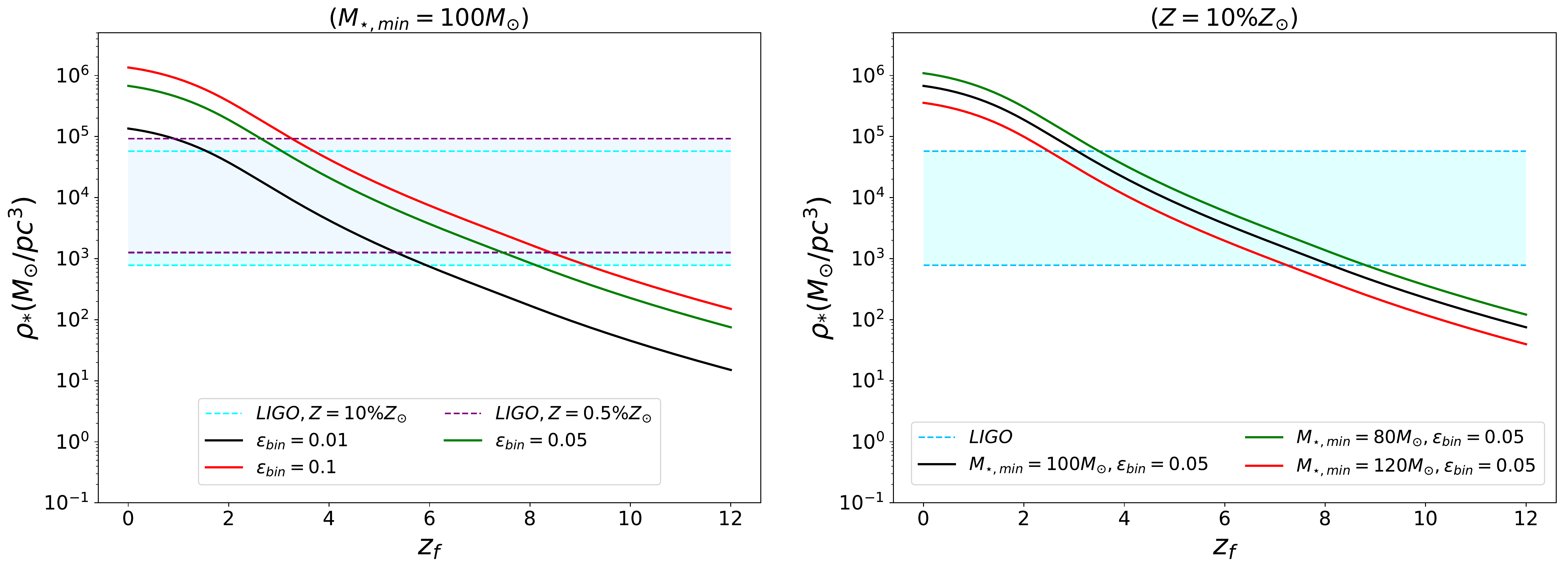}
	\caption{Comparison between the LIGO event density(shaded band), and the comoving star density $\rho_{\star}$. On the left panel, we choose  $M_{\star,min} = 100 M_{\odot}$ and show the star density for $\epsilon_{bin} = 0.01, 0.05, 0.1$, for two choices of metalliciy, $Z = 0.5 \% Z_{\odot} $ and $Z = 10 \% Z_{\odot} $. On the right panel, we show the case with $M_{\star, min} = 80, 100, 120 M_{\odot}$ and with $\epsilon_{bin} = 0.05$, and  $Z = 10 \% Z_{\odot} $.}
\label{SFRD100-10-N}
\end{figure*}

So far we have assumed that all of stars with mass in the range $50-150 M_{\odot}$ end up in a BBH. In reality, only a fraction of them collapse to BH in the desired mass range above $20 M_{\odot}$. This fraction depends on various parameters, such as the metallicity, magnetic field and rotation. In addition, not all of the generated BHs would end up in sufficiently tight BBHs that coalesce within $t_{H}$. We combine both of these factors through a parameter, $\epsilon_{bin}$. In principle, $\epsilon_{bin}$ could be time dependent, but for simplicity we take it to be a constant. Therefore, our derived limits on $\epsilon_{bin}$ should be taken as the constrains on the average of $\epsilon_{bin}$ accounting for all astrophysical channels.

Combining the different factors mentioned above, the resulting progenitor mass density is, 
\ba 
\label{SFD-Madau}
\rho_{\star}(z, M_{\star,min}) &=& f_{min}(M_{\star,min})\epsilon_{bin}  \int_{0}^{t(z)} \dot{\rho}_{\star} dt.
\ea
Figure \ref{SFRD100-10-N} shows $\rho_{\star}$ as a function of the formation redshift, hereafter $z_f$. On the left panel, we adopt $M_{\star,min} = 100 M_{\odot}$ and we show the star density for $\epsilon_{bin} = 0.01, 0.05, 0.1$.  Increasing $\epsilon_{bin}$ pushes us to higher redshift and so enhances the formation redshift. This makes sense as for higher efficiencies we increase the percentage of the star density for the BBH. We have also plotted the LIGO region for two different metalicites, $Z = 0.5 \% Z_{\odot} $ and $Z = 10 \% Z_{\odot} $. Interestingly, the metallicity does not affect the observational limit significantly implying that our results for the isolated stars are not strongly model dependent. This yields us  $  0.9 < z_f  <  9.13$ for the above range of parameters. 
In the right panel, we present the results for $M_{\star, min} = 80, 100, 120 M_{\odot}$. Here we have adopted $\epsilon_{bin} = 0.05$ and only present the LIGO results for $Z = 10 \% Z_{\odot}$. This shows that increasing $M_{\star, min}$ decreases $\rho_{\star}$ and also pushes towards lower redshifts. This makes sense since increasing  $M_{\star, min}$ we decreases $f_{min}$ in Eq.  (\ref{SFD-Madau}). This yields a redshift range  $ 3.3 < z_{f} < 8.94 $ for the above range of $M_{\star, min}$, including LIGO error bars.
\begin{figure}[!h] 
	\centering
	\includegraphics[width=0.48\textwidth]{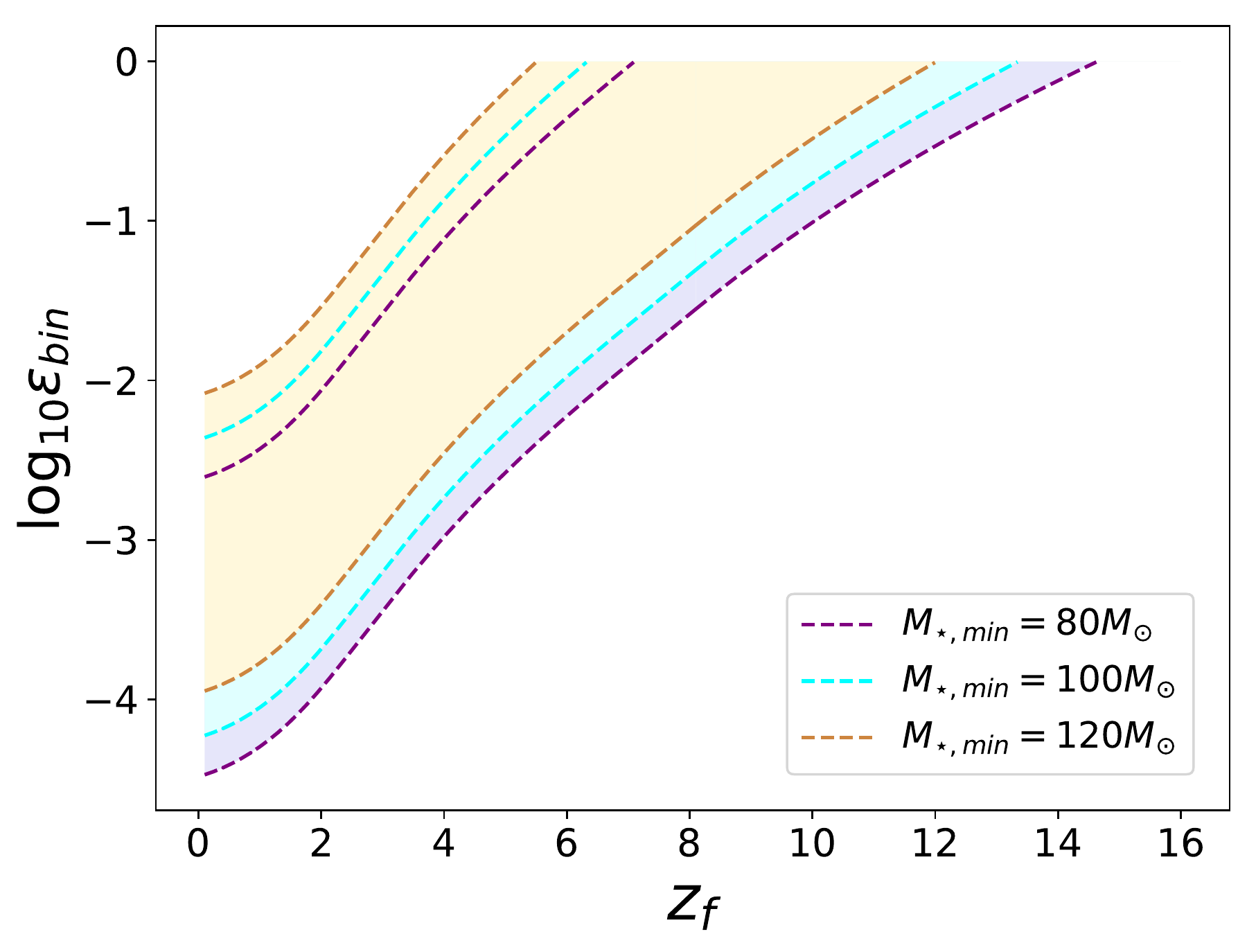}
	\caption{LIGO constraints on the progenitor mass fraction in coalescing BBH, \rm{$\log_{10}\epsilon_{bin}$}, as a function of formation redshift, $z_f$, for $M_{\star, min} = 80, 100, 120 M_{\odot}$.}
	\label{epsilon-z}
\end{figure}

Figure \ref{epsilon-z} presents the required progenitor mass fraction in coalescing binaries 
\rm{$\log_{10}\epsilon_{bin}$}, as a function of formation redshift, $z_{f}$, for
sourcing the LIGO event mass density for BHs with masses above $20 M_{\odot}$. Here we present the results for $M_{\star, min} = 80, 100, 120 M_{\odot}$. Again, increasing  $M_{\star, min}$ pushes toward lower redshifts.

In conclusion,  mapping the LIGO merger density to the star formation density we have found a range for the formation redshift of the LIGO progenitors. Our limit depends slightly on the threshold mass for the BHs progenitors. Assuming that all of the missing stars yield in the coalescing binaries requires their formation redshift to be $ 4.0 <z_f < 14 $. It is however more realistic to assume that only a fraction of the stars appear as binaries. If we assume  that $10 \%$ of the massive stars are in the form of the BBHs, we get 
$  2< z_f < 9 $.  Finally for the case with $1\%$  of stars in the for of binaries, the formation redshift is $  0 <z_f < 8$.
A late time suppression of $\epsilon_{bin}$ could result from the increase of metallicity in newly formed stars at low redshifts. 

While in this work we have considered a one-to-one mapping between the progenitor stars and the observed binaries, it would be interesting to generalize the current analysis to the case with non-zero delay time for the binary formation. We leave this investigation to a future paper.

We thank Daniel D'Orazio and John Forbes for helpful comments. We also very grateful to two anonymous referees for their constructive suggestions. R.E. acknowledges support by the Institute for Theory
and Computation at Harvard-Smithsonian Center for Astrophysics. This work was also supported in part by the Black Hole Initiative at Harvard University which is funded by a JTF grant.

\end{document}